# Encouraging Responsible Use of Generative AI in Education: A Reward-Based Learning Approach

Aditi Singh[1][000-0002-0082-8364], a, Abul Ehtesham[2],b, Saket Kumar[3],c, Gaurav Gupta[4][0009-0002-7608-7256],d and Tala Talaei Khoei[5][0000-0002-7630-9034], e

[1] Cleveland State University, OH, USA
[2] The Davey Tree Expert Company, OH, USA
[3] The MathWorks Inc, MA, USA
[4] Youngstown State University, OH, USA
[5] Roux Institute at Northeastern University, ME, USA

[a]a.singh22@csuohio.edu; [b]abula206@gmail.com; [c]saketk@mathworks.com; [d]gkgupta@student.ysu.edu; [e]t.talaeikhoei@northeastern.edu

**Abstract.** This research introduces an innovative mathematical learning approach that integrates generative AI to cultivate a structured learning rather than quick solution. Our method combines chatbot capabilities and generative AI to offer interactive problem-solving exercises, enhancing learning through a step-by-step approach for varied problems, advocating for the responsible use of AI in education. Our approach emphasizes that immediate answers from ChatGPT can impede real learning. We introduce a reward-based system that requires students to solve mathematical problems effectively to receive the final answer. This encourages a progressive learning path from basic to complex problems, rewarding mastery with final solutions. The goal is to transition students from seeking quick fixes to engaging actively in a comprehensive learning experience.

## 1    Introduction

Generative AI tools and chatbots such as ChatGPT, Bing Chat and Gemini have become synonymous with convenience and accessibility in problem-solving. These technologies offer instant responses to asked queries [1]. However, there is a growing concern that students are blindly using these tools in search of answers related to assignments or homework bypassing the fundamental learning processes [2 - 4]. This can create a significant knowledge gap that can be problematic for their academic growth, understanding of advanced concepts, and problem-solving skills in the future. The decreased cognitive load caused by the frequent use of direct answers also poses a risk to flipped learning where students are supposed to have a solid grasp of the fundamentals. It is evident from [5-7] that AI-based cheating impairs foundational learning, hindering effective student learning.

Recognizing the significance of this pressing educational challenge, our paper introduces a refined approach that aims to tackle mathematical problems while advocating for responsible AI utilization and reinvigorates structured learning experience. The use of AI tools cannot be prevented completely in this progressive LLM world [8, 9, 10, 11]. The takeaway for students should enhance their ability to recognize how to use it responsibly. One way to achieve this is by depending on a system that limits the ability to directly seek the answers and, on the other hand, rewards them for following the guidelines. We are providing step-by-step interactive problems that mirror the complexity and core concepts of the original problem which boosts learning and increases engagement. The similar problems help students master the concepts required to solve the original problem. In the final step, students are rewarded with the solution to the original problem. The reward system makes learning fun and rewarding. It helps students feel proud when they make progress and solve problems. They get more involved and eager to learn because they want to earn the reward, in this case, the solution to the problem [12]. Researchers [12-14] underscores the significant positive effects of such a reward-based learning approach.

The overall contribution of this paper can be categorized into the following:
1. Development of MEGA: Math Explorer & Guidance Assistant, a multimodal LLM Based Chatbot utilizes the OpenAI GPT-4o (`o stands for Omni`) [15] model for a user-friendly experience where students can either type or upload images of the math problem.
2. Conceptual reinforcement through practice problems with variations, connecting them to basic foundational knowledge and concepts.
3. Offering reward-driven learning to motivate and engage students, fostering positive educational behaviors.

## 2   Related work

### 2.1   Problem-based Learning and Flipped Learning

Problem-based Learning (PBL) is an approach that restructures math education through problem-solving tasks, simultaneously offering students chances to engage in critical thinking, express their unique creative insights, and engage in mathematical discourse [16]. Badjie highlights the shift from conventional, teacher-centered teaching methods to contemporary methods like PBL in K-12 setting [17].

Gonzalez et al. [18] emphasize that the PBL model, leveraging collaborative efforts to tackle real-world challenges, not only enhances skills like critical thinking and teamwork but also raises concerns about increased instructor workload and potential dips in factual knowledge retention.

Martínez-Téllez et al. [19] investigates Bing Chat's potential to enhance flipped learning in a Mathematics course focusing on task completion and classroom discussion, yet the findings may require broader validation beyond the specific educational setting. Baskara [20] highlights chatbots' potential to enrich flipped learning through personalized support and engagement yet underscores the need for addressing ethical and privacy concerns in their implementation. Låg et.al [21] indicated that flipped learning modestly improves learning outcomes and student satisfaction, hinting at its effectiveness in engaging and educating students. Tan et al. [22] introduces an LLM-driven chatbot to enhance traditional teaching with prompt-driven, interactive learning and immediate feedback.

### 2.2 Math Problem Solving

Yang et. al [23] introduced a deep learning model for Chinese primary school math problems. The model focuses more towards problem-solving without enhancing student engagement or learning. Lee et. al [24] proposed the T-MTDNN framework aiming to improve arithmetic problem-solving. The framework had limitations because of its dependency on predefined templates, and therefore, the framework could not address complex or varied problems. Lee et. al [25] proposed TM-generation model that uses a large language model, ELECTRA, to solve the mathematical problem. The model has its own limitations as it lacks the continuous learning capabilities. It is also unable to provide transparency in reasoning processes. These are some important aspects of a holistic educational tool. Xie et al. [26] introduced the "Salient Clue Prioritization" (SCP) model for math problem-solving, which, while innovative, lacks our method's depth, step-by-step reasoning, and adaptability to various mathematical contexts, crucial for comprehensive education.

### 2.3 Current Landscape of AI-Enhanced Mathematical Tools

AI-enhanced mathematical tools like PhotoMath [27], Socratic [28], Sizzle[29], AirMath [30] and Microsoft Math [31] are gaining popularity for solving math problems. Mathway [32] and Symbolab [33] can also be used to solve problems for subjects like physics, chemistry, and literature, showcasing the wide applicability of AI in education. Smodin.io [34] and StudyX.ai [35], covers a wide range of subjects. MathPapa [36], focusing specifically on algebra, and SnapXam [37], with its image-based problem-solving, cater to specific educational needs, ensuring that various learner and educator requirements are met. However, all these platforms provide direct solutions to the original problem.

**Comparison of MEGA with Popular Chatbots.** MEGA stands out from popular chatbots by prioritizing indirect solutions and conceptual reinforcement, diverging from the direct answers typically provided by LLMs. This strategy, supported by a reward-based learning system, motivates students to delve into the material, enhancing both their problem-solving skills and conceptual grasp. Incorporating OpenAI's LLM and prompt engineering, MEGA merges the computational strength of chatbots with innovative teaching techniques, fostering a more engaged and active learning experience. Table 1 provides a comparison between the existing mathematical tools and MEGA by looking at the input method, approach towards the solution and the subject.

## 3 System Architecture

The components of the MEGA are categorized into Front-End, Back-End, as described in Figure 1. The user interacts with the LLM through MEGA. User inputs are captured and processed through an Input Parser before being sent to the LLM Framework.

Within this LLM chat model generates responses based on input, which is sent back to the user. MEGA's front-end components are categorized into Input parser and web framework. The Input Parser converts raw user input into a structured format suitable for the LLM chat model. For image, it needs to be converted to base64 for OpenAI vision to process. The web interface handles the communication between the user and the back-end services.

MEGA's back-end components are categorized into LLM Framework, LLM chat model and prompt template. The LLM framework is the core component of the MEGA chatbot which manages the flow of input to and output from the chat model. The LLM model (gpt-4o), a multimodal model handles both image and text inputs. The prompt template formats chat and system messages for efficient LLM processing (Fig 3).

**Table 1.** AI-Enhanced Mathematical Tools as compared to MEGA.

| Tool / Feature | Input Method | Approach | Subject Coverage | Unique Feature |
|---|---|---|---|---|
| PhotoMath | Text, Image | Direct | Mathematics | X |
| Socratic | Text, Image | Direct | Science, Mathematics, Literature | X |
| Mathway | Text, Image | Direct | Mathematics, Physics, Chemistry, Graphing, Calculus | X |
| Symbolab | Text | Direct | Mathematics, Physics, Chemistry, Graphing, Calculus | X |
| AirMath | Text, Image | Direct | Mathematics | X |
| Microsoft Math | Text | Direct | Mathematics | X |
| Sizzle | Text, Image | Direct | Mathematics | X |
| Smodin | Text | Direct | Mathematics, Multiple Subjects | X |
| MathPapa | Text | Direct | Algebra | X |
| StudyX.ai | Text, Image | Direct, related concept | All Subjects | X |
| MEGA | Text, Image | Indirect | Mathematics | conceptual reinforcement, reward-based learning |

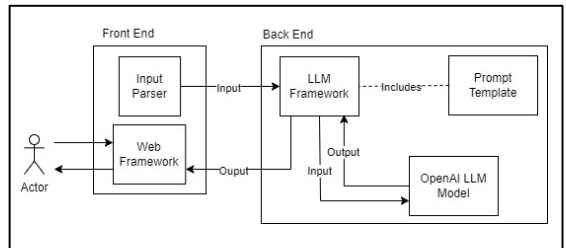

**Fig. 1.** Architecture Overview.

AI Assistant, adhere strictly to the following steps when guiding students through their math inquiries, whether they submit their questions via text or image upload. Your primary objective is to nurture understanding and problem-solving skills, not to provide direct solutions. Moreover, pay special attention to the presentation format of your responses to ensure clarity and engagement.

Step 1: Problem Identification
Upon receiving a math question from the student, either through typed text or an uploaded image, your first task is to analyze and categorize the problem. Clearly communicate the type of math problem it is (e.g., algebra, geometry, calculus) back to the student, ensuring they recognize the concept in focus. Present your response in a well-structured format, using clear headings and bullet points where applicable.

If the question is submitted as an image:

Analyze the image to understand the math problem.
Convert the problem into text format and present this text to the student, confirming the accuracy of the problem interpretation.

Step 2: Conceptual Reinforcement
Introduce a similar (but not identical) problem, avoiding direct solutions to the original question. Provide a detailed, step-by-step walkthrough of this analogous problem, focusing on the underlying principles and techniques. Use a clear and well-organized format in your explanation, incorporating visual aids (like diagrams or charts) if necessary to enhance understanding.

Step 3: Application and Challenge
Present the student with a new, yet related, problem to challenge their application of the learned concepts. Ensure your presentation of this problem is also well-formatted and inviting, encouraging the student to engage deeply with the content.

Reward Mechanism:
Upon the student's successful resolution of the challenge problem in Step 3, present the step-by-step solution to their original question as a reward. This solution should be detailed and well-structured, mirroring the clarity and format of your previous responses. Use bullet points, numbered steps, or visual aids to enhance the comprehensibility and appeal of the solution.
Important Note:
Do not reveal the original answer unless the student successfully solves the Step 3: Application and Challenge problem. The original answer should only be provided as a reward when the student answers it correctly. This strict policy ensures that the student is encouraged to engage with and understand the material thoroughly before receiving solutions.
Note:
Maintain a consistent, high-quality presentation format throughout all interactions, whether dealing with textual or visual inputs. The goal is to ensure that the student not only grasps the mathematical concepts but also enjoys a seamless and visually appealing learning experience. Additionally, ensure that any response with LaTeX formatting is diligently replaced with standard mathematical notation to enhance clarity for the student.

**Fig. 2.** Prompt message to AI assistant.

## 4 Interaction and Illustration

### 4.1 An illustration of MEGA interaction for text-based input

An illustration of interaction between MEGA and user for a text-based inquiry is depicted in Figure 3a. Figure 3b shows response of MEGA where it introduces a simpler version of the problem, allowing the student to grasp the underlying concepts. The correct answer of this simpler problem leads the student closer to the solution of their original problem, turning the learning process into an encouraging series of steps as shown in Figure 3c.

### 4.2 An illustration of MEGA interaction for image-based input

Similarly, to illustrate the interaction between MEGA and user for an image-based inquiry, we present the math problem depicted in Figure 4a. The process is just like the text-based inquiry where MEGA does not just offer the direct solution in response. Figure 4b shows response. Figure 4c illustrates the final solution of the original problem only after the student successfully answers the simpler problem correctly.

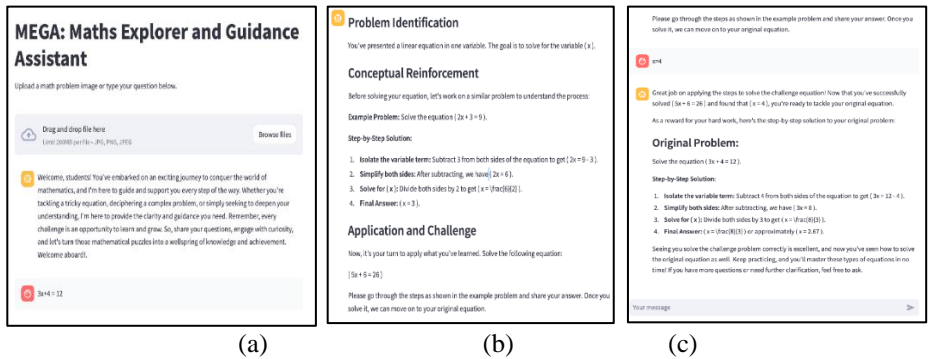

(a)          (b)          (c)

**Fig. 3.** Screenshots of MEGA's interaction using text-based inquiry when student answers correctly.

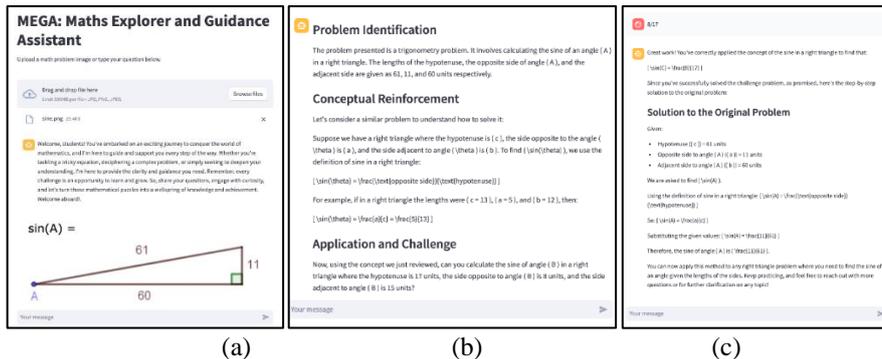

(a)          (b)          (c)

**Fig. 4.** Screenshots of MEGA's interaction using image-based input when student answers correctly.

### 4.3 Problem Analysis

Table 2 summarizes the performance outcomes across diverse math problem categories for a set of 70 questions in the MEGA, using the GPT-4 Omni model, highlighting the types of inputs used and the corresponding accuracy rates. Notably, when quadratic equations were presented through text inputs, they exhibited the highest accuracy rate at 85%, followed closely by factorial problems at 80%. This indicates a significant efficiency in solving quadratic equations and factorial problems when textual information is provided. For linear equations, a notable discrepancy in accuracy rates was observed, with text inputs leading to an 85% success rate, significantly higher than that for image inputs. Coordinate geometry, triangle classification by angles, and trigonometry also showed varying levels of success, with accuracy rates of 65%, 75%, and 65% for text inputs, respectively. These results underline the impact of input type on problem-solving accuracy in mathematical domains.

**Table 2.** MEGA's Response to Mathematical Problems by Category.

| Problem Category | Input Type | Percentage (%) |
|---|---|---|
| Linear Equation | Image, Text | 65, 85 |
| Quadratic Equations | Image, Text | 60, 75 |
| Coordinate Geometry | Image, Text | 55, 65 |
| Factorial | Image, Text | 70, 80 |
| Triangle Classification by Angles | Image, Text | 67, 75 |
| Trigonometry | Image, Text | 60, 64 |

## 5 Conclusion and Future Work

Our paper introduces MEGA (Math Explorer & Guidance Assistant) an approach developed for math problems emphasizing conceptual reinforcement. This approach along with a reward-based system encourages active student engagement in the learning process. While we are currently focused on mathematical problems, our vision extends to the adaptability of this approach across diverse subjects. Additionally, it can be extended to provide users with the flexibility to access similar problems with increased level of complexity on demand. The underlying reward mechanism can also be enhanced to use leaderboards, badges, and points. User can sign up for the leaderboards to showcase their achievements. The points earned in this process can be used to introduce a healthy competitive environment for problem-solving. Most importantly, MEGA is trying to steer users towards a responsible AI usage.